\documentclass[twocolumn,aps,pra,showpacs]{revtex4}

\usepackage{epsfig}

\newcommand{\bfk}{{\bf k}}
\newcommand{\bfK}{{\bf K}}

\newcommand{\ha}{\hat{a}}

\newcommand{\ddt}{\frac{d}{dt}}

\newcommand{\nl}{\nonumber\\}

\begin{document}


\title{Ultra-bright biphoton emission from an atomic vapor based on Doppler-free four-wave-mixing and collective emission}


\author{Y. P.  Huang}
\author{M. G. Moore}
\affiliation{Department of Physics \& Astronomy, Michigan State
University, East Lansing, MI  48824}


\date{\today}

\begin{abstract}
We propose a novel `butterfly' level scheme to generate highly
correlated photon pairs from atomic vapors. With multi-photon
Doppler-free pumping, background Rayleigh scattering is
dipole-forbidden and collective emission is permitted in all directions.
This results in usable pairs generated
simultaneously in the full $4\pi$ solid angle. Collecting these
pairs can produce photon pairs at a rate
of $\sim 10^{12}$ per second, given only moderate ensemble sizes of
$\sim 10^6$ atoms.
\end{abstract}


\maketitle

The study of correlated/entangled photon pairs has
long been a topic in the field of quantum optics
\cite{BouEkeZei00}. The importance of paired photons is two-fold:
they i) provide powerful tools to test the peculiar aspects of
quantum mechanics, such as violations of local-realism
 \cite{ScuZub97,GroPatKal07}; and  ii) they hold
promises for advancements in quantum measurement, communication, and
information processing \cite{BouPanMat97,MigDatSer98,
Luk03,FleImaMar05,SheKraOls06,VinMarPoo08,Kim08}. Over the past few
decades, spontaneous parametric down-conversion (SPDC) in nonlinear
crystals has been the standard source of photon pairs
\cite{HarOshBye67,KonEllFra05}. More recently, an alternative class
of biphoton sources has emerged, based on optical four-wave mixing
(FWM) in atomic vapors \cite{ZibLukScu99,VanEisAnd03,KuzBowBoo03,
BalBraKol05,KolDuBel06,ThoSimLoh06,ChaMatJen06,DuWenRu07,DuKolBel08}.
These approaches rely on collective effects \cite{Dic54} to greatly increase the
probability of correlated emission events.
Compared to SPDC, photon pairs generated via FWM in general have a
much narrower bandwidth, significantly greater temporal and spatial
coherence, and much higher conversion efficiencies. They are thus
particularly suitable for hybrid quantum communications and
computations employing atoms and photons \cite{DuaLukCir01,Kim08},
and for high-precision quantum measurements and imaging
\cite{MigDatSer98, VinMarPoo08}.

This far, proposed FWM photon pair sources can be
categorized into three types by level configuration. The first type,
built on atomic two-level systems, is a connected double-Rayleigh
emission process \cite{AspRogRey80,GraRigAso86}. Due to strong
background Rayleigh scattering, however, the
resulting pair correlation is very weak, without violating the
necessary Cauchy criteria for biphoton correlation \cite{KolDuBel06,
DuWenRu07}. A second type is configured on two-photon cascade
emission in a four-level system \cite{ChaMatJen06, ScuRay07}. While
high-fidelity photon pairs are generated, due to the unequal
wavelengthes of two cascade photons, the phase-matching condition
for collective emission can only be satisfied if the first photon is emitted by chance into a specific
small solid-angle, thus unpaired emission dominates the overall radiation, resulting in a relatively
low conversion efficiency.  The third type
employs Raman FWM (hereafter referred to as ``RFWM'') in multilevel
systems, configured on double-$\Lambda$
\cite{ZibLukScu99,BalBraKol05,KolDuBel06,DuWenRu07,DuKolBel08} or
``X'' \cite{ThoSimLoh06} level diagrams. The major challenge in
these schemes is to suppress background Rayleigh scattering, which
tends to rapidly overwhelm paired emission.
Three approaches have been proposed to overcome this
difficulty, including i) using frequency selectors to filter out
Rayleigh photons \cite{BalBraKol05}; ii) collecting pairs along
emission directions where the dipole pattern leads to zero Rayleigh
emission \cite{KolDuBel06}; and iii) using a single-mode optical
cavity to suppress Rayleigh transitions \cite{ThoSimLoh06}. While
yielding up to $10^5$ pairs per second, all of these setups are
unidirectional, where photon pairs are produced only along certain
directions. This restricts the obtainable beam brightness of the
photon pairs, since in each momentum mode, the time separation
between pairs must be sufficiently large to achieve strong
correlation effects. Lastly, in aforementioned FWM schemes where
unpaired emissions dominate, atomic samples are rapidly thermalized
due to random atomic recoils, limiting applications of these schemes
to ``hot'' vapors only.

To substantially increase the gain rate and suppress the atomic
thermalization, here we propose an omnidirectional biphoton source
configured on a `butterfly' level scheme. The central idea is to
completely eliminate the background Rayleigh scattering in RFWM by
employing electric-dipole forbidden driving channels. This can be
accomplished using multiphoton pumping. Combining this with a
Doppler-free geometry allows high-efficiency emission of photon
pairs in the full $4\pi$ solid angle. Collecting these pairs will
lead to twin beams of correlated or entangled photon pairs, whose
brightness can be tens of thousands times greater than that via
unidirectional schemes. Furthermore, since Rayleigh scattering has
been eliminated, the atomic thermalization will be strongly
suppressed, so that this scheme can also be applied to ultracold
vapors including Bose-Einstein condensates (BECs).

\begin{figure}
\epsfig{figure=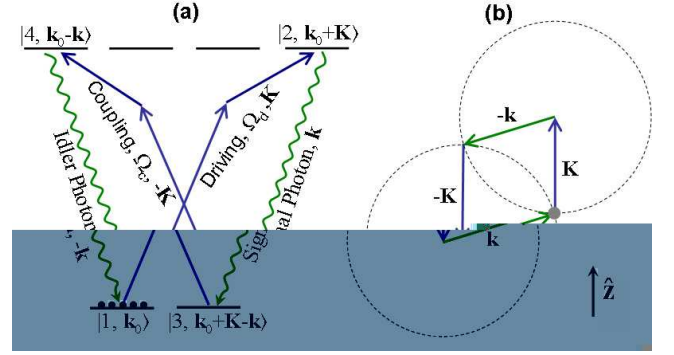, width=8.5 cm}
 \caption{(Color online) A schematic model of the butterfly scheme.
 Figure (a) draws the simplified level diagram, employing multi-photon
 driving and coupling pumps. The notation $|1,\bfk_0\rangle$
 indicates a single atom state in level $|1\rangle$ and with momentum $\hbar
 \bfk_0$, and so on. Figure (b) shows a closed,
  diamond-like dynamical cycle in atom-recoil momentum space,
  for an arbitrary $\bfk$ of signal
  photons.
 \label{fig1}}
\end{figure}

A schematic level diagram of the butterfly scheme is shown
in Fig. \ref{fig1} (a). While greatly simplified with respect to a
realistic level-scheme, this model will serve to illustrate the
important dynamical effects. An atomic ensemble, initially prepared
in the $|1\rangle$ state, is first weakly driven to the excited
$|2\rangle$ level via a multi-photon pump process, which imparts a net recoil momentum of $\hbar{\bf K}$, so that for an initial
momentum $\hbar\bfk_0$, an atom excited to $|2\rangle$ has a
momentum of $\hbar (\bfk_0+{\bf K})$. The atom then spontaneously
decays to $|3\rangle$, emitting a `signal' photon with a random momentum $\hbar \bfk$, shifting the atom's momentum to
$\hbar (\bfk_0+{\bf K}-\bfk)$. Decay to state $|1\rangle$, which would generate background Rayleigh scattering, is forbidden by dipole
selection-rules.
The atom in $|3\rangle$ state is then
rapidly repumped to $|4\rangle$ by a strong multi-photon coupling process. The
coupling is arranged to yield a net momentum of $-\hbar {\bf K}$,
leading to a momentum of $\hbar (\bfk_0-\bfk)$ for the atom. From $|4\rangle$,
the atom decays back to the $|1\rangle$ state, emitting an `idler' photon, which will be emitted with momentum $-\hbar\bfk$ due to collective enhancement.

The collective enhancement mechanism can be understood by noting first that the emission of the signal photon with momentum $\hbar\bfk$ imprints `which atom' information onto the atomic ensemble
via atomic recoil, provided of course that the single-photon recoil momentum is larger than the
momentum coherence length of the sample (which for a thermal gas of free particles is the inverse
sample length). If the idler photon is then emitted with momentum $-\hbar\bfk$, the atom is restored
to its initial momentum state of $\hbar\bfk_0$, thus `erasing' the `which atom' information, resulting in collective
emission by all atoms within a wavelength of the emission axis, enhancing the emission rate by a factor $\sim n\lambda^2R$,
where $n$ is the atomic density, $\lambda=2\pi/|\bfk|$, and $R$ is the cloud radius. The reason only this fraction of the atoms emit collectively is that the photons themselves carry sufficient information in their phase fronts to locate the transverse origin of emission with an accuracy given by the diffraction limit, $\lambda$. Conversely, emission of an `idler' photon with a momentum other than $-\hbar\bfk$ is not collective, and is therefore suppressed by unitarity.

The fact that the driving and coupling pumps
have opposite net momenta makes the scheme
`Doppler-free', so that phase-matched collective emission can occur regardless of which direction the signal photon randomly `chooses', resulting in omni-directional emission of correlated photon pairs. This is clearly seen in Fig.
\ref{fig1} (b), where for an arbitrary $\bfk$, the atomic dynamics
in the space of recoil momentum undergoes a closed, diamond-like
cycle.

To study the system's dynamics, we quantize the
electromagnetic field of signal and idler photons onto orthogonal
collective-emission modes $\{|\bfk\rangle\}$. Each mode subtends a solid angle $\sim (\lambda/R)^2$, centered on $\bfk$, and the modes do not overlap.
This quantization basis is mode matched to the collective emission angle. In this basis, the atomic dynamics is governed by a set of rate equations:
\begin{eqnarray}
\label{ndy}
    & & \!\!\!\!\!\!\!\!\!\! \ddt N_1 =
    \frac{i}{2}\left(\Omega_d\varrho_{21}-c.c\right)+\sum_\bfk
    \Gamma_4 \mu_{\bfk 4} N_{\bfk 4} (N_1+1), \\
    & & \!\!\!\!\!\!\!\!\!\! \ddt N_2 =
    -\frac{i}{2}\left(\Omega_d\varrho_{21}-c.c\right)-\sum_\bfk
    \Gamma_2 \mu_{\bfk 2} N_2 (N_{\bfk 3}+1), \\
    & & \!\!\!\!\!\!\!\!\!\! \ddt \varrho_{21}=i \frac{\Omega_d}{2} (N_1-N_2) \nl
    & & ~~~~~~+\frac{1}{2}\varrho_{21}
      \sum_\bfk\left[\Gamma_4\mu_{\bfk 4} N_{\bfk 4}-
       \Gamma_2 \mu_{\bfk 2} (N_{\bfk 3}+1) \right], \\
    & & \!\!\!\!\!\!\!\!\!\! \ddt N_{\bfk 3} = \frac{i}{2} \left(\Omega_c\varrho_{\bfk 43}-c.c\right)+
    \Gamma_2 \mu_{\bfk 2} N_2 (N_{\bfk 3}+1), \\
    & & \!\!\!\!\!\!\!\!\!\! \ddt N_{\bfk 4} =
    -\frac{i}{2}\left(\Omega_c \varrho_{\bfk 43}-c.c\right)-
    \Gamma_4  N_{\bfk 4} (\mu_{\bfk 4}N_1+1), \\
     & & \!\!\!\!\!\!\!\!\!\! \ddt \varrho_{\bfk 43}=i \frac{\Omega_c}{2} (N_{\bfk 3}-N_{\bfk 4}) \nl
    & & ~~~~~~+\frac{1}{2}\varrho_{\bfk 43}
      \left[\Gamma_2\mu_{\bfk 2} N_2-
       \Gamma_4(\mu_{\bfk 4} N_1+1) \right].
\end{eqnarray}
Here, $N_1$ and $N_2$ are the atom numbers in states $|1\rangle$
and $|2\rangle$, and $\varrho_{21}$ is the corresponding
coherence term. $N_{\bfk 3}$ is the number of atoms in $|3\rangle$, whose signal-photon recoil kick corresponds to emission into collective mode $|\bfk\rangle$.
Similarly, $N_{\bfk 4}$ is the number of these atoms transferred to the state $|4\rangle$ by the coupling laser and $\rho_{\bfk 4 3}$ is the coherence between the two
states. In the above equations, $\Gamma_2, \Gamma_4$ are the
spontaneous emission rates for $|2\rangle \rightarrow |3\rangle$ and
$|4\rangle\rightarrow |1\rangle$ decays, while $\mu_{\bfk 2}$ and
$\mu_{\bfk 4}$ are the collective enhancement factors, which have been taken to be real. The
imaginary part of $\mu_{\bfk 2}, \mu_{\bfk, 4}$, describing the
laser-induced dipole-dipole interaction, is in general orders of
magnitude smaller than the real part, and can thus be neglected.
For a spherical sample of radius $R$, we find $\mu_{\bfk
j}=(1-|\hat{k}\cdot\hat{d}_j|^2)\frac{3}{8\pi}\left(\frac{\lambda}{2R}\right)^2$.
Here, $\hat{d}_2, \hat{d}_4$ are the unit vector
directed along the dipole moments of the $|2\rangle \rightarrow
|3\rangle$ and $|4\rangle\rightarrow |1\rangle$ transitions.

The total emission rates for signal and idler photons,
corresponding to the (enhanced) decay rates of the $|2\rangle$ and
$|4\rangle$ levels, are $R_S=\Gamma_2 \sum_\bfk \mu_{\bfk 2}
N_2(N_{\bfk 3}+1)$ and $R_I=\Gamma_4 \sum_\bfk \mu_{\bfk 4} N_{\bfk
4}(N_1+1)$, respectively. Assuming steady-state, clearly we must
have $R_I\le R_S$. If $R_I<R_S$, more signal photons are generated
than idler photons, so that pairing is weak. Thus, at a minimum,
strong pairing requires $R_I=R_S$. Focusing on a single $\hat{k}$
direction, and assuming $\mu_{\bfk 2}=\mu_{\bfk 4}$ and
$\Gamma_2=\Gamma_4$, we find
$\frac{R_I(\hat{k})}{R_S(\hat{k})}=\left(\frac{N_1+1}{N_2}\right)\left(\frac{N_{\bfk
4}}{N_{\bfk 3}+1}\right)$. For a strong drive, $\Omega_d\gtrsim \Gamma_2$, we have $N_2\sim
N_1$, so that we require $N_{\bfk 4}\sim N_{\bfk 3}\gg 1$,
which thus requires a strong coupler, $\Omega_c\gtrsim \Gamma_4\mu_{\bfk 4}N_1$, as well. We find that dynamically this
doesn't work, instead population builds up in $N_{\bfk 3}$ and
$N_{\bfk 4}$ without strong pairing. This leaves the case of weak
driving, $\Omega_d\ll \Gamma_2$ so that $N_2\ll N_1$. This then requires $N_{\bfk
4}=\frac{N_2(N_{\bfk 3}+1)}{N_1}$, which can be arranged by
adjusting the drive and coupler strengths and detunings, and for
$N_{\bfk 4}\ll 1$ results in strong pairing.

The total number of photon pairs generated
per second is given by $\Gamma_4 \sum_\bfk \mu_{\bfk 4}
N_{\bfk 4}(N_1+1)$. For a BEC, the  condensate atom loss rate, due to
spontaneous emission from the $|4\rangle$ level, is
$\Gamma_4 \sum_\bfk N_{\bfk 4}$. The ratio of pair generation and atom
loss rates is $\kappa=N_1\bar{\mu}$, where $\bar{\mu}=\sum_\bfk \mu_{\bfk
4} N_{\bfk 4}/\sum_\bfk N_{\bfk 4}$. For a spherical sample this is
roughly $\lambda^2/4\pi R^2$. Typically, $\kappa$ is much greater than one, so that many
pairs can be generated before rogue photon emission destroys the BEC. This means
that if desired, a BEC can act as an ultra-bright photon-pair source without being destroyed in the process. We note that in general, an
atom which emits spontaneously from $|4\rangle$ receives an additional photon recoil-kick, but is still fully able to participate in collective
emission of subsequent photon pairs. In a more realistic level scheme this may not be true for atoms which spontaneously decay from $|4\rangle$ to a
level other than $|1\rangle$.

To measure the time correlation of the two photons, we calculate the
time-averaged second-order correlation function
$g^{(2)}(\bfk,-\bfk,\tau)$ \cite{ScuZub97},
\begin{eqnarray}
    g^{(2)}(\bfk,-\bfk,\tau)\!\!=\!\!\frac{1}{T}\!\!\!\int_0^T \!\! \!\! dt
     \frac{\langle \ha^\dag_{\bfk s}(t)\ha^\dag_{-\bfk i} (t\!\!+\!\!\tau)
     \ha_{-\bfk i}(t\!\!+\!\!\tau)\ha_{\bfk s}(t)\rangle}
     {\langle \ha^\dag_{\bfk s}(t)\ha_{\bfk s}(t)\rangle\langle
     \ha^\dag_{-\bfk i}(t\!\!+\!\!\tau)\ha_{-\bfk i}(t\!\! +\!\!
     \tau)\rangle}, \nonumber
\end{eqnarray}
where $\ha_{\bfk s}$ and $\ha_{\bfk i}$ are the annihilation
operators for  signal and idler photons, which may or may not differ in polarization for a given $\bfk$, and $T$ is
the averaging window. This is evaluated by first using adiabatic
following to write the photon operators in terms of the atomic
operators. The resulting expressions are then factorized into
products of the occupation numbers. The equations for the occupation
numbers are solved in steady-state with atom losses neglected, which
allows us to take $T\rightarrow \infty$. In this manner, we find to
a good approximation that, $
    g^{(2)}(\bfk,-\bfk, \tau)\approx \!\! 1+ \!\!\frac{1} {N_{\bfk 4}}\chi(\tau).
$
For a strong coupling with $\Omega_c\gtrsim \Gamma_4 \mu_{\bfk
4}N_1$, we find $\chi(\tau)\approx \sin^2\left(
\frac{1}{2}\Omega_{c}\tau\right) \exp(-\frac{1}{2}\Gamma_4\mu_{\bfk
4} N_1\tau)$. The correlation then exhibits
oscillatory and damped behavior with sharp peaks, indicating strong
time-correlation between the signal and idler photons
\cite{BalBraKol05}. The time delay of idler photons is roughly
$(\Gamma_4\mu_\bfk  N_1)^{-1}$, which is much shorter than the
temporal coherence length of the signal photons. This ensures strong
interferences of paired photons, for example, when mixed in a
polarized beamsplitter \cite{ThoSimLoh06}.

\begin{figure}
\epsfig{figure=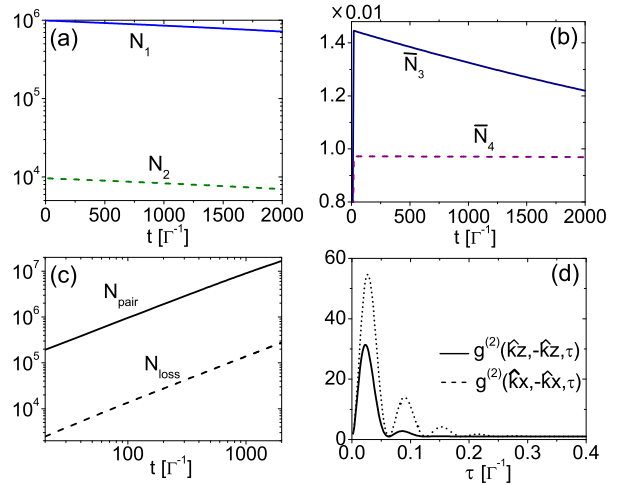, width=9.0 cm}
 \caption{(Color online) Figures (a)-(b) show the evolutions of
 $N_1,N_2,\overline{N}_3, \overline{N}_{4}$, while (c) compares the number of generated
 photon pairs $N_{pair}$ and lost atoms $N_{loss}$.
 Figure (d) plots the second-order correlation function
 $g^{(2)}(\bfk,-\bfk,\tau)$ for $\bfk=k\hat{z}$ and $\bfk=k\hat{x}$,
 respectively. Parameters are given in text.
 \label{fig2}}
\end{figure}

As an example, we solve the rate-equation dynamics for a symmetric
Gaussian sample of $N=10^6$ atoms. We consider resonant driving
and coupling pumps propagating along $\hat{z}$ and $-\hat{z}$
directions, and assume $\Gamma_2=\Gamma_4\equiv\Gamma$. We choose
$\Omega_d=0.1 \Gamma$, $\Omega_c=100\Gamma$ and $R=50\lambda$. The results are
shown in Fig. \ref{fig2} (a)-(d). In figure (a), we plot the time
evolutions of $N_1, N_2$, showing that at all times only a small
fraction of atoms are excited, as required. In figure (b), we plot
the average occupation numbers $\overline{N}_{3}$ and
$\overline{N}_{4}$, obtained by averaging over $\bfk$.  Both are
found to be of order of $0.01$, so that there is negligible overlap between subsequent pairs in a given mode $\bfk$.
The total photon-pair number and lost atom number are shown
in figure (c), where they are found to increase linearly in time
with fitted rates of $8.3\times 10^3 \Gamma$ and $1.4 \times 10^2
\Gamma$. In figure (d), we plot the second-order correlation functions for photon pairs propagating
along $\pm\hat{z}$ and $\pm\hat{x}$ directions. Both cases exhibit sharp peaks of widths $\sim
0.05 \Gamma^{-1}$, indicating strong temporal correlation which violate the standard
Cauchy-Schwartz inequality by a factor $\gtrsim 1000$.

We now examine the polarization entanglement of paired photons.
For $\hat{d}_{2,4}=\frac{1}{\sqrt{2}}(\hat{x}\pm i\hat{y})$, the
probabilities for signal photons to be left(right) circularly
polarized along $\bfk$, denoted
$\hat{\epsilon}_L(\hat{\epsilon}_R)$, are
$\beta^{L}_{S}(\theta)=\left(1+\cot^4 \frac{\theta}{2}\right)^{-1}$
and $\beta^R_{S}(\theta)=\beta^L_{S}(\pi-\theta)$, with $\theta$
defined relative to $\bfK$. Similarly, for the idler photons we
have
 $\beta^R_{I}=\beta^L_S
(\theta)$, and $\beta^L_{I}=\beta^R_S (\theta)$. The probability for
photons to be in opposite circular polarizations along $\bfk$
(thus in the same polarizations along each's propagating direction)
is $
    P(\theta)=(1+\cot^8\frac{\theta}{2})/\left(1+\cot^4
    \frac{\theta}{2}\right)^2,
$ which is extremely flat around $\theta=0,\pi$, where
$P(\theta)\approx 1-\frac{1}{8}\textmd{mod}(\theta,\pi)^4\approx 1$,
meaning that photon pairs emitted over a wide range of $\theta$ will
yield strong polarization entanglement. Due to the
temporal overlap of signal and idler photons, each pair emitted
within the strong correlation angle is approximately in the Bell state of
$|\Psi^+\rangle=\frac{1}{\sqrt{2}}
(|\epsilon_R\epsilon_L\rangle+|\epsilon_L\epsilon_R\rangle)$. For
example, pairs with one photon emitted within $\theta<0.5$
(corresponding to $17\%$ of the emitted pairs) have an
entanglement fidelity $\ge99\%$.
\begin{figure}
\epsfig{figure=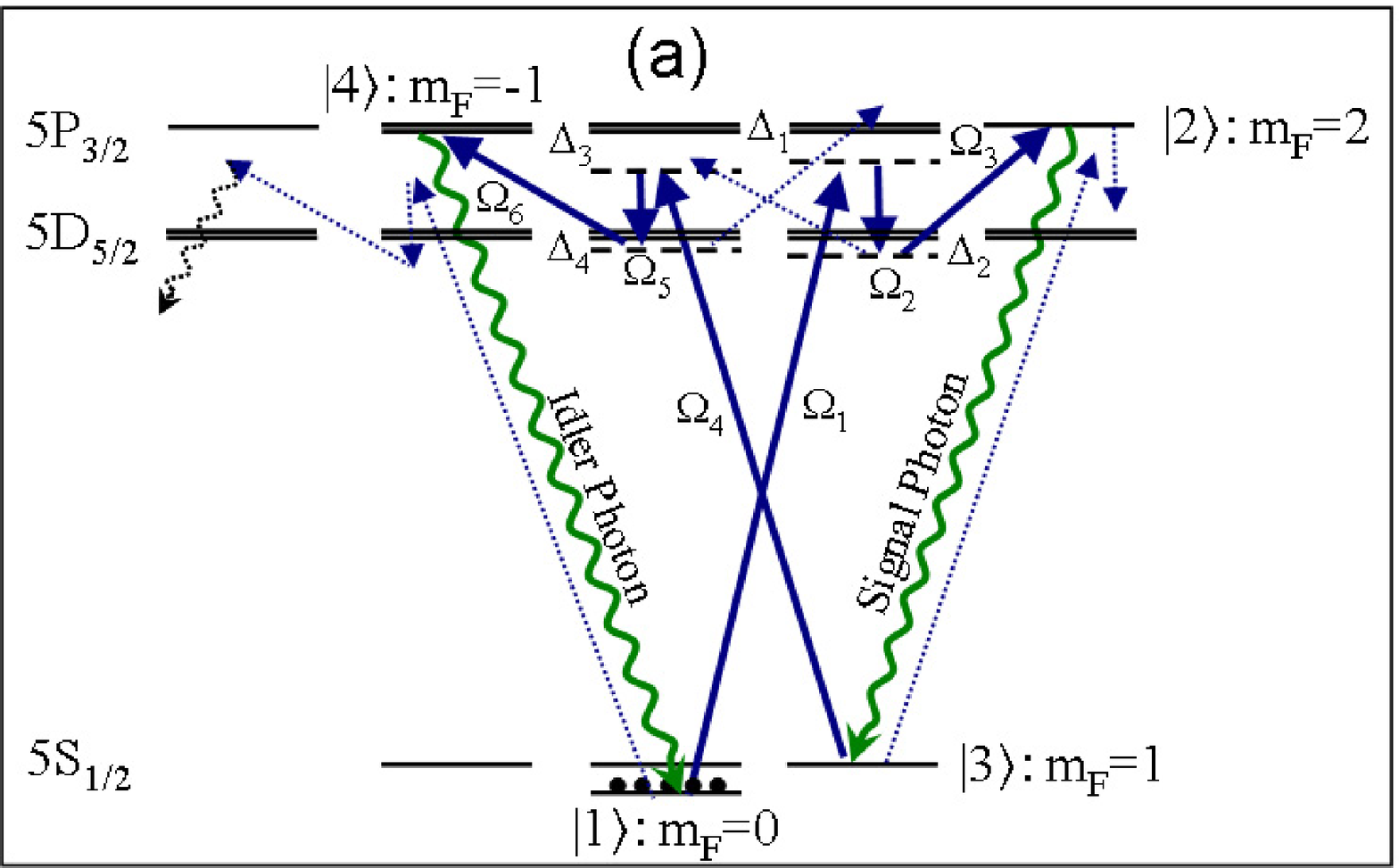, width=8.0 cm} \epsfig{figure=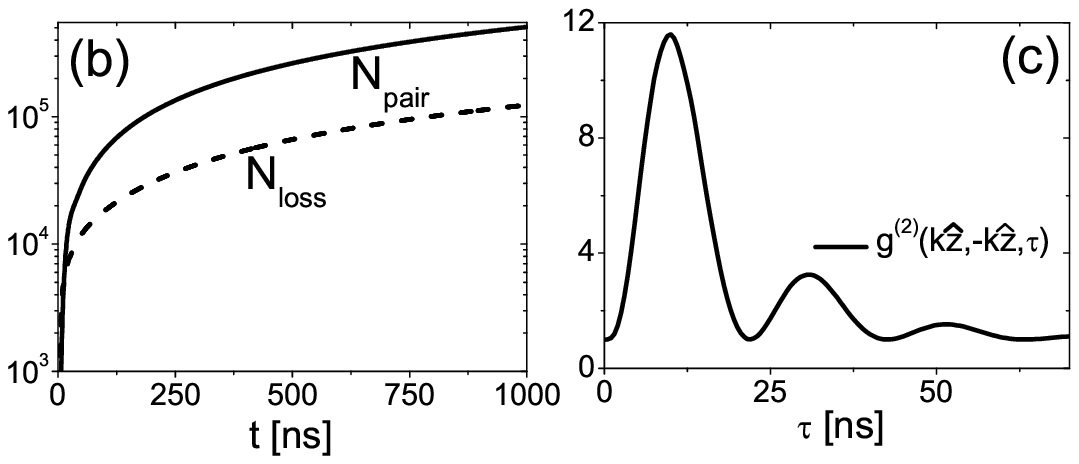,
width=9.0 cm}
    \caption{(Color online) A realistic butterfly scheme using Ag
    atoms. Figure (a) shows the level diagram, where thicker solid lines
    draw dominating coupling channels, while thinner dashed lines
    show weaker side transitions. Figure (b) plots the time evolutions of
    $N_{pair}$ and $N_{loss}$, while (c) draws the time correlation function
    for photon pairs emitted along $\pm \hat{z}$ directions.
 \label{fig3}}
\end{figure}

We now consider a realistic butterfly level scheme configured on the
$328$nm-line of the $5^2 S _{1/2}\leftrightarrow 5^2P_{3/2}$
transition in Silver, as shown in figure \ref{fig3} (a). The atoms
are prepared in the $|1\rangle\equiv|F=0,m_F=0\rangle $ state and then
follow a FWM cycle which deposits them in level $|3\rangle$. A second independent FWM cycle then returns them
to the initial $|1\rangle$ state. The drive (coupling) FWM cycle consists of one UV laser,and two infrared lasers, with Rabi frequencies $\Omega_{1(4)}$ $\Omega_{2(5)}$ and
$\Omega_{3(6)}$, respectively.
Signal photons are
emitted as $|2\rangle$ atoms spontaneously decay to the sole
dipole-allowed state of $|3\rangle\equiv |F=1,m_F=1\rangle$, and idler
photons are generated as each atom in $|4\rangle$ (actually 2 effectively degenerate hyperfine levels) decays collectively back to
$|1\rangle$, and to its initial momentum $\hbar\bfk_0$ (atoms which spontaneously decay to other hyperfine levels are `lost' as they can no longer participate in the collective emission cycle). We note that the frequencies of the signal and idler photons
differ by the ground hyperfine splitting, which is
smaller than the superradiance-broadened linewidth of the idler
photons, thus guaranteeing the frequency indistinguishability of the
photons.

For a spherical cloud of $10^6$ atoms with a radius of $R=20\mu
m$, and with (all in units of GHz): $\Omega_1=0.4$, $\Omega_2=9$,
$\Omega_3=1.5$, $\Omega_4=4$, $\Omega_5=12$, $\Omega_6=0.5$,
$\Delta_1=24$, $\Delta_2=3$, $\Delta_3=80$, $\Delta_4=0.4$, we solve
the rate equations numerically. Results are shown in figure
\ref{fig3} (b) and (c). In figure (b), the production rate of photon
pairs is $\sim 0.5\times 10^{12}$ per second. The ratio of generated pairs
$N_{pair}$ to lost atoms $N_{loss}$ is about $4$, so that roughly
$10^6$ pairs of photons can be generated before significant atomic
loss. In figure (c), the time correlation function
exhibits sharp peaks, thus showing strong pair correlation between
signal and idler photons.

The dominant loss mechanism is due to atoms from $|1\rangle$ being pumped to $|4\rangle$ by $\Omega_4$, and then decaying non-collectively. Due the the detuning of this mechanism, the rogue photons are different in frequency from signal and idler
photons by $\sim 80$ GHz, which can be filtered out.
When the atom loss becomes significant, the system will enter a more complicated regime, where the system tries to equilibrate,
resulting in macroscopic occupation of all ground hyperfine sub-levels, and presumably diminished pair correlations. If this is the case, then only a small number of photon pairs will be generated, but in a very short burst. In this case, this particular system may be an excellent source for generating highly number-difference squeezed twin beams for quantum interferometry.

Acknowledgement: we thank A. Leanhardt for helpful discussions. This
work is supported in part by National Science Foundation Grant No.
PHY0653373.

\bibliographystyle{apsrev}
\bibliography{biblio}

\end{document}